\documentclass[conference]{IEEEtran}
\IEEEoverridecommandlockouts
\usepackage{amsmath}
\usepackage{amssymb}
\usepackage{amsfonts}
\usepackage{algorithm}
\usepackage{dsfont}
\usepackage{float}
\usepackage{cite}
\usepackage{graphicx}
\usepackage{epsfig}
\usepackage{subfigure}
\usepackage{psfrag}
\usepackage{xcolor}
\usepackage{url}
\usepackage[colorlinks,linkcolor=black,urlcolor=black,anchorcolor=black,citecolor=black,hyperfootnotes=true]{hyperref}

\newtheorem{thm}{\textbf{\textup{Theorem}}}

\newtheorem{Prop}{Proposition}

\newcommand{\tabincell}[2]{\begin{tabular}{@{}#1@{}}#2\end{tabular}}

\begin{document}
	\title{
		Detection of Abrupt Change in Channel Covariance Matrix for Multi-Antenna Communication
	}
	\author{\IEEEauthorblockN{Runnan Liu${}^\dagger{}^\ddagger$, Liang Liu${}^\ddagger$, Dazhi He${}^\dagger$, Wenjun Zhang${}^\dagger$, and Erik G. Larsson${}^\ast$}
		\IEEEauthorblockA{${}^\dagger$ SEIEE, Shanghai Jiao Tong University. Email: \{liurunnan,hedazhi,zhangwenjun\}@sjtu.edu.cn\\ ${}^\ddagger$ Department of EIE, The Hong Kong Polytechnic University. Email: liang-eie.liu@polyu.edu.hk \\ ${}^\ast$ Department of Electrical Engineering, Link\"oping University, Email: erik.g.larsson@liu.se}
	}
	\maketitle

	\begin{abstract}
		The knowledge of channel covariance matrices is of paramount importance to the estimation of instantaneous channels and the design of beamforming vectors in multi-antenna systems. In practice, an abrupt change in channel covariance matrices may occur due to the change in the environment and the user location. Although several works have proposed efficient algorithms to estimate the channel covariance matrices after any change occurs, how to detect such a change accurately and quickly is still an open problem in the literature. In this paper, we focus on channel covariance change detection between a multi-antenna base station (BS) and a single-antenna user equipment (UE). To provide theoretical performance limit, we first propose a genie-aided change detector based on the log-likelihood ratio (LLR) test assuming the channel covariance matrix after change is known, and characterize the corresponding missed detection and false alarm probabilities. Then, this paper considers the practical case where the channel covariance matrix after change is unknown. The maximum likelihood (ML) estimation technique is used to predict the covariance matrix based on the received pilot signals over a certain number of coherence blocks, building upon which the LLR-based change detector is employed. Numerical results show that our proposed scheme can detect the change with low error probability even when the number of channel samples is small such that the estimation of the covariance matrix is not that accurate. This result verifies the possibility to detect the channel covariance change both accurately and quickly in practice.
	\end{abstract}


	\section{Introduction}
	In modern wireless systems, multiple-input multiple-output (MIMO) is a core technology to improve the channel capacity and reliability. However, the MIMO beamforming design crucially relies on the knowledge of the channel state information. Along this line, many efficient algorithms have been proposed to estimate the user instantaneous channels \cite{Hassibi03,Gershman06}.

In practice, user channels are spatially correlated due to the dependent antenna patterns at the base station (BS) and the finite number of scatters \cite{Gao15,Sanguinetti20}. In this case, the acquisition of the channel covariance matrices is important as well. On one hand, to estimate the instantaneous channels, we need to estimate the channel covariance matrices first, because the minimum mean-squared error (MMSE) channel estimators are usually functions of the channel covariance matrices \cite{Marzetta16}. On the other hand, there is a new trend to design the beamforming vectors based on the channel covariance matrices \cite{Park17,Jin19,Zhao21}, instead of the instantaneous channels. This is because channel covariance matrices change much more slowly than instantaneous channels, and the overhead to estimate these matrices is thus much lower. Such a property is especially appealing for downlink massive MIMO systems \cite{Marzetta16,marzetta2010noncooperative,larsson2014massive} and intelligent reflecting surface (IRS) systems \cite{Liaskos08,Renzo19,Zhang21}, where the overhead for estimating so many channel coefficients in each coherence block is huge \cite{bjornson2016massive,wang2020channel}. Motivated by the above reasons, several interesting works have been done on channel covariance matrix estimation \cite{Chin,upadhya2018covariance}.

Despite slowly, the channel covariance matrices do vary in practice due to the change in the environment and the user location. A key question is when an abrupt change occurs, how to detect it accurately and quickly such that we can re-estimate the channel covariance matrices as soon as possible. In this paper, we focus on this challenge in a point-to-point communication system, where a multi-antenna BS serves a single-antenna user equipment (UE). A novel change detection protocol is proposed as shown in Fig. 1, where a proper number of coherence blocks form a frame, and the pilot signals received across a small number of blocks at the beginning of each frame will be used to detect whether the channel covariance matrix changes over the previous frame. The complicated approaches proposed in \cite{Chin,upadhya2018covariance} will be only adopted to estimate the channel covariance matrix in the subsequent blocks when a change is detected.

Under this protocol, an intuitive way for change detection is to first estimate the new covariance matrix very accurately and then compare it with the previous one. However, a precise estimation usually requires pilot signals over a sufficiently long time, making quick change detection impossible. In this paper, we apply the classic change detection theory \cite{basseville1993detection} to tackle this issue. First, we propose a genie-aided change detector based on the log-likelihood (LLR) test, where the covariance matrix after change is assumed to be known. The probabilities of missed detection and false alarm are characterized, which can sever as performance limit. Then, we consider the practical case without knowledge of the covariance matrix in the new frame. Maximum likelihood (ML) technique is used to estimate the new covariance matrix, and the LLR-based change detector is then proposed based on the estimated covariance matrix. Numerical results show that our proposed scheme can achieve very low probabilities of missed detection and false alarm even when the number of detection blocks is small, because a moderate (instead of exact) estimation of the covariance matrix is a sufficient evidence under our proposed change detector.

\section{System Model}\label{Sec_SysMod}
	\begin{figure}
		\setlength{\abovecaptionskip}{-0.1 cm}
		\centering
		\includegraphics[width=9 cm]{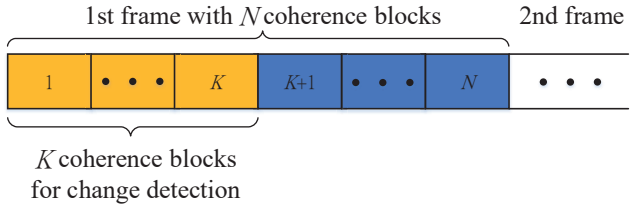}
		\caption{Proposed change detection protocol: each frame consists of $N$ coherence blocks, the first $K$ of which are used to detect whether the channel covariance matrix has changed compared to that in the previous frame.  }
		\label{SysMod}
	\end{figure}

	Consider a narrow-band communication system where a BS equipped with $M$ antennas serves a single-antenna UE. We assume a block fading channel model, where channel stays constant in one coherence block, but may vary independently over different blocks. Let $\boldsymbol{h}\in \mathbb{C}^{M\times 1}$ denote the channel between the BS and the UE. In this paper, we consider a Rayleigh fading channel model, i.e., $\boldsymbol{h}\sim \mathcal{CN}(\boldsymbol{0},\boldsymbol{C}^h)$, where $\boldsymbol{C}^h$ denotes the covariance matrix of $\boldsymbol{h}$. In practice, $\boldsymbol{C}^h$ changes much more slowly than $\boldsymbol{h}$. Our aim is to detect the change in $\boldsymbol{C}^h$ accurately and quickly.

	To realize this goal, in this paper, we propose a novel change detection protocol, as shown in Fig. \ref{SysMod}, which can be applied in both downlink and uplink communication. First, we define a frame as the collection of $N$ coherence blocks. In practice, we can measure the average frequency for the occurrence of the change in $\boldsymbol{C}^h$ and set $N$ properly based on this frequency such that $\boldsymbol{C}^h$ will change at most once over one frame. Then, based on the pilot signals received at the first $K$ coherence blocks of each frame, the system needs to determine whether the channel covariance matrix in the current frame, denoted by $\boldsymbol{C}^h_1$, is the same as that in the previous frame, denoted by $\boldsymbol{C}_0^h$. In other words, we need to detect between two hypothesis
	\begin{equation}
		\begin{aligned}
			H_0:& \boldsymbol{C}^h=\boldsymbol{C}^h_1=\boldsymbol{C}^h_0,\\
			H_1:& \boldsymbol{C}^h=\boldsymbol{C}^h_1\neq \boldsymbol{C}^h_0.
		\end{aligned}
	\end{equation}If a change has been detected, i.e., $H_1$ is true, we can use more coherence blocks following the first $K$ coherence blocks to estimate the new channel covariance matrix $\boldsymbol{C}^h_1$ based on the existing techniques \cite{5484583,Chin,upadhya2018covariance}. Otherwise, if $H_0$ is true, the estimated covariance matrix remains the same as that in the previous frame. As a result, at any frame, we always know the channel covariance matrix in the previous frame. In the rest of this paper, we focus on how to perform change detection based on the pilot signals received at the first $K$ coherence blocks of each frame, with the prior knowledge of channel covariance matrix in the previous frame.

	Specifically, at the $k$-th coherence block of a frame, $1\leq k\leq K$, the received pilot signals in the uplink and downlink are respectively give as
	\begin{align}
& \boldsymbol{Y}^{{\rm ul}}_k=\sqrt{\rho}\boldsymbol{h}_k\boldsymbol{x}^H + \boldsymbol{N}^{{\rm ul}}_k, \label{Eq_SysRxSignal1} \\
&			\boldsymbol{y}^{{\rm dl}}_k = \sqrt{\rho}\boldsymbol{X}\boldsymbol{h}_k+\boldsymbol{n}^{{\rm dl}}_k, ~ k=1,\ldots,K, \label{Eq_SysRxSignal}
    \end{align}where $\boldsymbol{h}_k\backsim \mathcal{CN}(\boldsymbol{0},\boldsymbol{C}^h_1)$ denotes the channel at the $k$-th block of the considered frame, $\rho$ denotes the transmit power, $\boldsymbol{x}\in \mathbb{C}^{T\times 1}$ and $\boldsymbol{X}\in \mathbb{C}^{T\times M}$ denote the uplink and downlink pilot signals with length $T$, and $\boldsymbol{N}^{{\rm ul}}_k\in\mathbb{C}^{M\times T}\backsim \mathcal{CN}(\boldsymbol{0},\sigma^2 \boldsymbol{I})$ and $\boldsymbol{n}^{{\rm dl}}_k\backsim \mathcal{CN}(\boldsymbol{0},\sigma^2 \boldsymbol{I})$ denote the additive white Gaussian noise (AWGN) at the BS side and the UE side. In the uplink, the pilot sequence satisfies $\boldsymbol{x}^H\boldsymbol{x}=T$, while in the downlink, we assume that the pilot sequence length satisfies $T\geq M$ such that $\boldsymbol{X}^H\boldsymbol{X}=T\boldsymbol{I}$.

	One intuitive way to perform change detection is to first estimate $\boldsymbol{C}_1^h$ very accurately and then detect whether a change has occurred in $\boldsymbol{C}_0^h$. However, an accurate estimation of $\boldsymbol{C}_1^h$ depends on: 1. an accurate estimation of $\boldsymbol{h}_k$ at each coherence block via some advanced signal processing technique; and 2. a sufficiently large number of samples about $\boldsymbol{h}_k$'s. However, advanced channel estimation techniques, e.g., MMSE estimation, usually require the channel statistical information, which is not available here because we do not know whether the channel covariance matrix is still $\boldsymbol{C}_0^h$. Moreover, in practice, any change should be detected within a short time duration, and an accurate estimation of $\boldsymbol{C}_1^h$ over a longer time duration is only performed after (not before) a change is detected.
	
	To address the above issues under the protocol shown in Fig. \ref{SysMod}, this paper proposes the following change detection scheme. First, the channel at each coherence block of the phase detection phase is estimated based on the ML technique\footnote{After the change detection phase in the frame, Bayesian channel estimation techniques, e.g., MMSE estimation, can be applied based on the new channel statistics to replace the ML estimation for reducing estimation mean-squared error.}, which requires no statistical information of the channel. For the received pilot signals (\ref{Eq_SysRxSignal1}) and (\ref{Eq_SysRxSignal}), the ML channel estimators at coherence block $k$ in the uplink and downlink can be unified as
	\begin{equation}
			\tilde{\boldsymbol{h}}_k
			=\boldsymbol{h}_k + \tilde{\boldsymbol{n}}_k,~ k=1,\ldots,K,
		\label{tildey}
	\end{equation}
	where $\tilde{\boldsymbol{n}}_k\backsim \mathcal{CN}(\boldsymbol{0},\frac{\sigma^2}{E_0}\boldsymbol{I})$ with $E_0=\rho T$.  Second, after the ML channel estimators in the $K$ coherence blocks, denoted by  $\tilde{\boldsymbol{H}}=[\tilde{\boldsymbol{h}}_1,...,\tilde{\boldsymbol{h}}_K]\in \mathbb{C}^{T\times K}$, are obtained, we aim to apply the LLR test based on the change detection theory \cite{basseville1993detection} such that a change can be detected accurately just utilizing a small number of channel samples.

	To introduce the change detection theory \cite{basseville1993detection} and provide performance upper bound, in the following, we first focus on how to perform LLR test in the ideal case that if there is a change of the channel covariance matrix in the current frame, the new covariance matrix is a known matrix $\boldsymbol{C}_1^h\neq \boldsymbol{C}_0^h$. Then, we will focus on LLR test for change detection in the practical scenario where the channel covariance matrix in the current frame $\boldsymbol{C}_1^h$ is unknown and needs to be estimated roughly first.
	
	\section{Change Detection with Known $\boldsymbol{C}^h_1$}
	\subsection{LLR-based Covariance Matrix Change Detector}
	In this section, we propose a genie-aided covariance matrix change detector, where if a change occurs, the new covariance matrix $\boldsymbol{C}_1^h$ is known with the help of some genie. Our detector is based on the standard LLR test. Specifically, for a given $\boldsymbol{C}^h_i, i=0,1$, the conditional probability density function (PDF) of the ML estimator $\tilde{\boldsymbol{h}}_k$ is
	\begin{equation}
		\begin{aligned}
			\begin{small}
				p\big(\tilde{\boldsymbol{h}}_k|\boldsymbol{C}^h_i\big) = \frac{1}{\sqrt{(2\pi)^T|\boldsymbol{C}_i^{\tilde{H}}|}}\exp\bigg[-\frac{1}{2}\tilde{\boldsymbol{h}}^H_k\big(\boldsymbol{C}_i^{\tilde{H}}\big)^{-1}\tilde{\boldsymbol{h}}_k\bigg],
			\end{small}
		\end{aligned}
		\label{Prob_i}
	\end{equation}
	where $\boldsymbol{C}_i^{\tilde{H}}$ is the covariance matrix of $\tilde{\boldsymbol{h}}_k$ given $\boldsymbol{C}^h_i$, i.e.,
	\begin{equation}
		\begin{aligned}
			\boldsymbol{C}_i^{\tilde{H}} =  \boldsymbol{C}^h_i+\frac{\sigma^2}{E_0}\boldsymbol{I}, ~ i=0,1.
		\end{aligned}
		\label{Eq_CovMTXCY}
	\end{equation}
	Then, the LLR at block $k$ for change detector is defined as
	\begin{equation}
		\begin{small}
			LLR_k(\tilde{\boldsymbol{h}}_k,\boldsymbol{C}_0^h,\boldsymbol{C}_1^h)\triangleq\log\Big(p\big(\tilde{\boldsymbol{h}}_k|\boldsymbol{C}^h_1\big)\Big)-\log\Big(p\big(\tilde{\boldsymbol{h}}_k|\boldsymbol{C}^h_0\big)\Big).
		\end{small}
		\label{LLR}
	\end{equation}
	Moreover, define the sum of the LLR over all the $K$ blocks for change detection as
	\begin{equation}
		\begin{aligned}
			S(\tilde{\boldsymbol{H}},\boldsymbol{C}^h_0,\boldsymbol{C}^h_1)=&\sum_{k=1}^{K}LLR_k(\tilde{\boldsymbol{h}}_k,\boldsymbol{C}_0^h,\boldsymbol{C}_1^h)\\
			=&
			\tilde{S}(\tilde{\boldsymbol{H}},\boldsymbol{C}^h_1)-\tilde{S}(\tilde{\boldsymbol{H}},\boldsymbol{C}^h_0),
		\end{aligned}
		\label{S}
	\end{equation}
	where $\tilde{S}(\tilde{\boldsymbol{H}},\boldsymbol{C}^h_i)$ is defined as
	\begin{equation}
		\begin{aligned}
			\tilde{S}(\tilde{\boldsymbol{H}},\boldsymbol{C}^h_i) =&\sum_{k=1}^{K}\log\Big(p\big(\tilde{\boldsymbol{h}}_k|\boldsymbol{C}^h_i\big)\Big)\\
			=&-\frac{K}{2}\bigg(T\log2\pi+\log\Big(|\boldsymbol{C}^h_i+\frac{\sigma^2}{E_0}\boldsymbol{I}|\Big) \\ &+\mathrm{tr}\Big(\big(\boldsymbol{C}^h_i+\frac{\sigma^2}{E_0}\boldsymbol{I}\big)^{-1}\boldsymbol{S}_{\tilde{H}}\Big)\bigg),
		\end{aligned}
		\label{S1}
	\end{equation}
	with
	\begin{equation}
		\boldsymbol{S}_{\tilde{H}}=\frac{1}{ K}\tilde{\boldsymbol{H}}\tilde{\boldsymbol{H}}^H=\frac{\sum_{k=1}^K \tilde{\boldsymbol{h}}_k \tilde{\boldsymbol{h}}_k^H}{K}
		\label{SampleCov}
	\end{equation}
	denoting the sample covariance matrix of the ML channel estimators. Then, the LLR-based change detector given the genie-aided knowledge of $\boldsymbol{C}_1^h$ is defined as \cite{basseville1993detection}
	\begin{equation}
		S(\tilde{\boldsymbol{H}},\boldsymbol{C}^h_0,\boldsymbol{C}^h_1)\mathop{\lessgtr}^{H_0}_{H_1} \Theta,
		\label{criterion}
	\end{equation}
	where $\Theta$ is a threshold.
	
	\subsection{Detection Error Analysis}\label{DetErrAna}
	In this subsection, we analyze the missed detection probability and the false alarm probability under our proposed genie-aided change detector. Particularly, missed detection means that $H_1$ is true, but the change detector claims that $H_0$ is true, while false alarm means that $H_0$ is true, but the change detector claims that $H_1$ is true. The probabilities of miss detection and false alarm can thus be expressed as
	\begin{equation}
		\begin{aligned}
			P_{MD} &= P\big(S(\tilde{\boldsymbol{H}},\boldsymbol{C}^h_0,\boldsymbol{C}^h_1)\leq\Theta|H_1\big),\\
			P_{FL} &= P\big(S(\tilde{\boldsymbol{H}},\boldsymbol{C}^h_0,\boldsymbol{C}^h_1)\geq\Theta|H_0\big).
		\end{aligned}
		\label{DetectErr}
	\end{equation}
	
	\begin{thm}\label{thm}
		Define
		\begin{equation}
			\boldsymbol{q}_i=\textup{eig}\bigg(\Big(\big(\boldsymbol{C}_i^h+\frac{\sigma^2}{E_0}\boldsymbol{I}\big)^{\frac{1}{2}}\Big)^H\boldsymbol{M}\big(\boldsymbol{C}_i^h+\frac{\sigma^2}{E_0}\boldsymbol{I}\big)^{\frac{1}{2}}\bigg),\quad i=0,1,
			\label{q}
		\end{equation}
		where $\textup{eig}(\cdot)$ denotes the vector consisting of the eigenvalues of a matrix, and
		\begin{equation}
			\boldsymbol{M} = \big(\boldsymbol{C}_0^h+\frac{\sigma^2}{E_0}\boldsymbol{I}\big)^{-1}-\big(\boldsymbol{C}_1^h+\frac{\sigma^2}{E_0}\boldsymbol{I}\big)^{-1}.
			\label{Mdef}
		\end{equation}
		Moreover, define
		\begin{equation}
			R =\log\Big(\big|\boldsymbol{C}_0^h+\frac{\sigma^2}{E_0}\boldsymbol{I}\big|\Big)-\log\Big(\big|\boldsymbol{C}_1^h+\frac{\sigma^2}{E_0}\boldsymbol{I}\big|\Big).
			\label{R}
		\end{equation}
		Then, the probabilities of missed detection and false alarm under the genie-aided change detector (\ref{criterion}) are given by
		\begin{equation}
			\begin{aligned}
				P_{MD}&= p\Big(\xi(\boldsymbol{q}_1,K,M)<2(\Theta-KR)\Big),\\
				P_{FA}&=  p\Big(\xi(\boldsymbol{q}_0,K,M)<-2(\Theta-KR)\Big),
			\end{aligned}
			\label{AnalysisResults}
		\end{equation}
		where $\xi(\boldsymbol{q}_i,K,M)=\sum_{m=1}^{M}q_{i,m}\chi^2_{2K,m}$ obeys generalized chi-squared distribution \cite{das2020method}, with $q_{i,m}$ denoting the $m$-th element of $\boldsymbol{q}_i$ and $\chi^2_{2K,m}$'s representing independent chi-squared distribution variables with $2K$ degree of freedom.
	\end{thm}

\begin{IEEEproof}
Please refer to Appendix \ref{appendix1}.
\end{IEEEproof}

	The probabilities of missed detection and false alarm in (\ref{AnalysisResults}) are expressed as cumulative distribution function (CDF) of $\xi(\boldsymbol{q}_i,K,M)$, whose close-form cannot be easily expressed \cite{davies1973numerical,das2020method}. However, we can apply the characteristic-function-based method in \cite{davies1973numerical} to the approximation of the CDF of $\xi(\boldsymbol{q}_i,K,M)$ shown in (\ref{AnalysisResults}) quickly and accurately.
	
	\section{Change Detection with Unknown $\boldsymbol{C}^h_1$}\label{Sec_ML}
	In this section, we consider the practical scenario where $\boldsymbol{C}_1^h$ is unknown. Different from the ideal case that $\boldsymbol{C}_1^h$ is known, we will first estimate $\boldsymbol{C}_1^h$ roughly based on $\tilde{\boldsymbol{H}}$ (\ref{tildey}) by using the ML technique, based on which the LLR-based detector is then used to detect whether a change occurs.
	
	Specifically, according to the change detector (\ref{criterion}), the ML estimator of $\boldsymbol{C}_1^h$ given $\tilde{\boldsymbol{H}}$ is defined as the covariance matrix that can maximize the sum of the log-likelihood over the $K$ coherence blocks, i.e.,

	\begin{equation}
		\hat{\boldsymbol{C}}_1^h = \arg \max_{\boldsymbol{C}_1^h} \tilde{S}(\tilde{\boldsymbol{H}},\boldsymbol{C}_1^h),
		\label{ML_est}
	\end{equation}
	where $\tilde{S}(\tilde{\boldsymbol{H}},\boldsymbol{C}_1^h)$ is given in (\ref{S1}). As shown in \cite{anderson1970estimation}, the optimal solution to (\ref{ML_est}) is $\hat{\boldsymbol{C}}_1^h=\boldsymbol{S}_{\tilde{H}}-\frac{\sigma^2}{E_0}\boldsymbol{I}$, where $\boldsymbol{S}_{\tilde{H}}$ is the sample covariance matrix given in (\ref{SampleCov}).

	However, when $K\ll M$, the ML estimator based on the sample covariance matrix is ill-conditioned. To improve the estimation performance, we introduce new constraints to problem (\ref{ML_est}) for reducing the condition number of the estimated covariance matrix. Specifically, the modified problem to obtain the ML estimator of $\boldsymbol{C}_1^h$ is given by
	\begin{equation}
		\mathcal{P}_1 \left\{
		\begin{aligned}
			\mathop{\textup{Maximize}}_{\boldsymbol{C}^h_1}&\quad \tilde{S}(\tilde{\boldsymbol{H}},\boldsymbol{C}_1^h)\\
			\textup{Subject to}&\quad \beta\boldsymbol{I}\preceq\boldsymbol{C}^h_1\preceq \kappa\beta\boldsymbol{I},
		\end{aligned}
		\right.
		\label{CostFun1}
	\end{equation}where $\beta>0$ and $\kappa\beta$ with $\kappa>1$ are the lower bound and upper bound for the minimum eigenvalue and the maximum eigenvalue, respectively, to ensure that the estimated covariance matrix is well-conditioned. It is non-trivial to solve Problem $\mathcal{P}_1$, because it is non-convex. However, in the following, we manage to characterize its closed-form solution.
	
	\begin{Prop}\label{Prop1}
		Consider the following problem
		\begin{equation}
			\mathcal{P}_2\left\{
			\begin{aligned}
				\mathop{\textup{Minimize}}_{\boldsymbol{C}_1^{\tilde{H}}}&\quad \log|\boldsymbol{C}_1^{\tilde{H}}|+\mathrm{tr}\big(({\boldsymbol{C}_1^{\tilde{H}}})^{-1}\boldsymbol{S}_{\tilde{H}}\big)\\
				\textup{Subject to}&\quad ( \beta+\frac{\sigma^2}{E_0})\boldsymbol{I}\preceq\boldsymbol{C}_1^{\tilde{H}} \preceq	( \kappa\beta+\frac{\sigma^2}{E_0})\boldsymbol{I}.
			\end{aligned}
			\right.
			\label{CostFun1_1}
		\end{equation}
		
		Let $\hat{\boldsymbol{C}}^{\tilde{H}}$ denote the optimal solution to Problem $\mathcal{P}_2$. Then, the optimal solution to Problem $\mathcal{P}_1$ is

		\begin{equation}
			\hat{\boldsymbol{C}}_1^h = \hat{\boldsymbol{C}}^{\tilde{H}} - \frac{\sigma^2}{E_0}\boldsymbol{I}.
			\label{ChCY}
		\end{equation}
	\end{Prop}
	
	\begin{IEEEproof}
		We prove Proposition \ref{Prop1} by contradiction. Suppose that the optimal solution to Problem $\mathcal{P}_1$ is not given by (\ref{ChCY}). Let $\tilde{\boldsymbol{C}}_1^h\neq \hat{\boldsymbol{C}}_1^h$ denote this optimal solution. Then, consider a new solution to Problem $\mathcal{P}_2$ as $\tilde{\boldsymbol{C}}^{\tilde{H}}=\tilde{\boldsymbol{C}}_1^h+\frac{\sigma^2}{E_0}\boldsymbol{I}$. Since $\tilde{\boldsymbol{C}}_1^h \succeq \beta\boldsymbol{I}$ and $\tilde{\boldsymbol{C}}_1^h \preceq \kappa\beta\boldsymbol{I}$, $\tilde{\boldsymbol{C}}^{\tilde{H}}$ is a feasible solution to Problem $\mathcal{P}_2$. Moreover, $\log|\tilde{\boldsymbol{C}}^{\tilde{H}}|+\mathrm{tr}\big(({\tilde{\boldsymbol{C}}^{\tilde{H}}})^{-1}\boldsymbol{S}_{\tilde{H}}\big)<\log|\hat{\boldsymbol{C}}^{\tilde{H}}|+\mathrm{tr}\big(({\hat{\boldsymbol{C}}^{\tilde{H}}})^{-1}\boldsymbol{S}_{\tilde{H}}\big)$ since $\hat{\boldsymbol{C}}_1^h$ is a feasible solution to Problem $\mathcal{P}_1$ and thus $\tilde{S}(\tilde{\boldsymbol{H}},\tilde{\boldsymbol{C}}_1^h)>\tilde{S}(\tilde{\boldsymbol{H}},\hat{\boldsymbol{C}}_1^h)$. This contradicts to the fact that $\hat{\boldsymbol{C}}^{\tilde{H}}$ is the optimal solution to Problem $\mathcal{P}_2$. Proposition \ref{Prop1} is thus proved.
	\end{IEEEproof}
	According to Proposition 1, we can solve Problem $\mathcal{P}_1$ via solving Problem $\mathcal{P}_2$. To get the optimal solution to Problem $\mathcal{P}_2$, we first define
	\begin{equation}
		\boldsymbol{C}^R = \big(\boldsymbol{C}_1^{\tilde{H}}\big)^{-1}.
		\label{Inv}
	\end{equation}
	Then, it can be shown that Problem $\mathcal{P}_2$ is equivalent to the following problem.
	\begin{equation}
		\mathcal{P}_3\left\{
		\begin{aligned}
			\mathop{\textup{Minimize}}_{\boldsymbol{C}^R}& \quad -\log\big(|\boldsymbol{C}^R|\big)+\mathrm{tr}\big(\boldsymbol{C}^R\boldsymbol{S}_{\tilde{H}}\big)\\
			\textup{Subject to}&\quad \frac{E_0}{E_0\kappa\beta+\sigma^2}\boldsymbol{I}\preceq \boldsymbol{C}^R \preceq \frac{E_0}{E_0\beta+\sigma^2}\boldsymbol{I},\\
		\end{aligned}
		\right.
		\label{CostFun3}
	\end{equation}
	
	In the following, we show that there exists a closed-form solution to Problem $\mathcal{P}_3$. First, consider the objective function of Problem $\mathcal{P}_3$. Define the eigenvalue decomposition of $\boldsymbol{C}^R$ and $\boldsymbol{S}_{\tilde{\boldsymbol{H}}}$ as
	\begin{equation}
		\begin{aligned}
			\boldsymbol{C}^R&=\boldsymbol{\Phi}_C\boldsymbol{\Lambda}_C\boldsymbol{\Phi}_C^H,\\
			\boldsymbol{S}_{\tilde{H}}&=\boldsymbol{\Phi}_S\boldsymbol{\Lambda}_S\boldsymbol{\Phi}_S^H,
		\end{aligned}
	\end{equation}
	where $\boldsymbol{\Phi}_C$ and $\boldsymbol{\Phi}_S$ consist of the eigenvectors of $\boldsymbol{C}^R$ and $\boldsymbol{S}_{\tilde{H}}$, and $\boldsymbol{\Lambda}_C=\textup{diag}(\boldsymbol{\lambda}_C)$ and $\boldsymbol{\Lambda}_S=\textup{diag}(\boldsymbol{\lambda}_S)$ with $\boldsymbol{\lambda}_C=[\lambda_{C,1},\ldots,\lambda_{C,M}]^T$ and $\boldsymbol{\lambda}_S=[\lambda_{S,1},\ldots,\lambda_{S,M}]^T$ consisting of the eigenvalues of $\boldsymbol{C}^R$ and $\boldsymbol{S}_{\tilde{H}}$. It then follows that
	\begin{subequations}
		\begin{align} &-\log\big(|\boldsymbol{C}^R|\big)+\mathrm{tr}\big(\boldsymbol{C}^R\boldsymbol{S}_{\tilde{H}}\big)\\
			=&-\log\big(|\boldsymbol{\Lambda}_C|\big)+\mathrm{tr}\big(\boldsymbol{\Phi}_C\boldsymbol{\Lambda}_C\boldsymbol{\Phi}_C^H\boldsymbol{\Phi}_S\boldsymbol{\Lambda}_S\boldsymbol{\Phi}_S^H\big)\\
			\geq & -\log\big(|\boldsymbol{\Lambda}_C|\big)+\mathrm{tr}\big(\boldsymbol{\Lambda}_C\boldsymbol{\Lambda}_S\big)\label{Ineq}\\
			=&\sum_{m=1}^{M} \big(-\log\lambda_{C,m}+\lambda_{C,m}\lambda_{S,m}\big),
		\end{align}
	\end{subequations}
	where the equality in (\ref{Ineq}) holds when $\boldsymbol{\Phi}_C=\boldsymbol{\Phi}_S$. In other words, the minimum of the objective function is achieved when the eigenvectors of $\boldsymbol{C}^R$ are the same as those of $\boldsymbol{S}_{\tilde{H}}$, i.e., $\boldsymbol{\Phi}_C=\boldsymbol{\Phi}_S$. Moreover, the constraints in Problem $\mathcal{P}_3$ are on the eigenvalues of $\boldsymbol{C}^R$, not its eigenvectors. As a result, under the optimal solution to Problem $\mathcal{P}_3$, it must follow that

	\begin{equation}
		\boldsymbol{\Phi}_C=\boldsymbol{\Phi}_S.
	\end{equation}
	We can take this property into Problem $\mathcal{P}_3$ such that only the eigenvalues of $\boldsymbol{C}^R$ need to be optimized. Then, Problem $\mathcal{P}_3$ reduces to
		\begin{equation}
			\mathcal{P}_4\left\{
			\begin{aligned}
				\mathop{\textup{Minimize}}_{\{\lambda_{C,m}\}}& \sum_{m=1}^{M} \big(-\log\lambda_{C,m}+\lambda_{C,m}\lambda_{S,m}\big)\\
				\textup{Subject to} &\quad \frac{E_0}{E_0\kappa\beta+\sigma^2}\leq\lambda_{C,m}\leq \frac{E_0}{E_0\beta+\sigma^2}, ~ \forall m.
			\end{aligned}
			\right.
		\end{equation}

	It can be shown that the optimal solution to Problem $\mathcal{P}_4$ is
	\begin{equation}
		\hat{\lambda}_{C,m} = \min\bigg(\max\Big(\frac{E_0}{E_0\kappa\beta+\sigma^2},\frac{1}{\lambda_{S,m}}\Big),\frac{E_0}{E_0   \beta+\sigma^2}\bigg).
		\label{Optimial_lamda}
	\end{equation}
	Define $\hat{\boldsymbol{\lambda}}_C=[\hat{\lambda}_{C,1},\ldots,\hat{\lambda}_{C,M}]^T$. Then, the optimal solution to Problem $\mathcal{P}_3$ is
		\begin{equation}
			\hat{\boldsymbol{C}}^R=\boldsymbol{\Phi}_S \textup{diag}(\hat{\boldsymbol{\lambda}}_C)\boldsymbol{\Phi}_S^H.
		\end{equation}
	Then, according to Proposition \ref{Prop1} and (\ref{Inv}), we have the following Theorem.
	
	\begin{thm}
		The optimal solution to Problem $\mathcal{P}_1$ is
		\begin{equation}
			\hat{\boldsymbol{C}}^h_{\textup{ML}}=  \boldsymbol{\Phi}_S (\textup{diag}(\hat{\boldsymbol{\lambda}}_C))^{-1}\boldsymbol{\Phi}_S^H-\frac{\sigma^2}{E_0}\boldsymbol{I},
			\label{ML}
		\end{equation}
		where $\boldsymbol{\Phi}_S$ consists of the eigenvectors of the sample covariance matrix $\boldsymbol{S}_{\tilde{H}}$ given in (\ref{SampleCov}), and the elements in $\hat{\boldsymbol{\lambda}}_C=[\hat{\lambda}_{C,1},\ldots,\hat{\lambda}_{C,M}]^T$ are defined by (\ref{Optimial_lamda}).
	\end{thm}
	
	At last, with the ML estimation of $\boldsymbol{C}_1^h$ given in (\ref{ML}), the change detector (\ref{criterion}) will reduce to

	\begin{equation}
		S(\tilde{\boldsymbol{H}},\boldsymbol{C}^h_0,\hat{\boldsymbol{C}}_{\textup{ML}}^h)\mathop{\lessgtr}^{H_0}_{H_1} \Theta.
		\label{Criterion_ML}
	\end{equation}

\begin{figure*}[!t]
		\normalsize
		\begin{equation}
			\boldsymbol{C}^h_{m1,m2} = \frac{1}{2\pi}\int_{0}^{2\pi}\exp\bigg[-j\frac{2\pi}{\alpha}D_{m1,m2}\sin\Omega\Big(1-\frac{\Psi^2}{4}+\frac{\Psi^2\cos(2\theta)}{4}\Big)+\Psi D_{m1,m2}\cos\Omega\sin\theta\bigg] d\theta, ~ \forall m1, m2.
			\label{ChannelMod}
		\end{equation}
		\hrulefill
	\end{figure*}
	
	\section{Numerical Results}
	In this section, we present numerical results to verify the effectiveness of the proposed change detectors. Due to the space limitations, we just consider downlink in the simulation. All relevant simulation parameters are given in Table \ref{ParaTable}. Specifically, the BS has $M=32$ antennas, while the pilot sequence length in the considered $K$ coherence blocks is set to be $T=M=32$. The pilot sequence $\boldsymbol{X}$ is generated based on the discrete Fourier transform (DFT) matrix.  The carrier frequency is selected as a typical E-band mmWave frequency at 80$\mathrm{GHz}$. We set the signal-to-noise ratio (SNR) for the communication between the BS and the UE as $0 \textup{dB}$. Moreover, we use the ``one-ring'' model to describe the covariance matrix of the UE channel, which is shown in (\ref{ChannelMod}) at the top of next page. Particularly, $\boldsymbol{C}_{m1,m2}^h$ denotes the element at the $m1$-th row and $m2$-th column of $\boldsymbol{C}^h$, $\Omega$ denotes the angle of departure (AoD), $\Psi$ denotes the angle spread, $\alpha$ denotes the wavelength of the signal, $D_{m1,m2}=2(m1-m2)\alpha$ is the relative spacing distance between antenna $m1$ and antenna $m2$, and $\theta$ denotes the angle of a possible scatter around the UE. In the rest of this section, we assume that the change in the channel covariance matrix is caused by the change of AoD. We use $\Delta\Omega$ to denote the change of AoD.


	\begin{table}[t]
		\vspace{0cm}
		\centering
		\caption{Simulation parameters}
		\label{ParaTable}
		\vspace{-0.1cm}
		\begin{tabular}{lcc}
			\setlength{\abovecaptionskip}{0.cm}
			\setlength{\belowcaptionskip}{-1.cm}
			\\[-2mm]
			\hline
			\hline\\[-2mm]
			{\bf \small Parameter}\qquad & {\bf\small Notation}\qquad & {\bf\small Value}\\
			\hline
			\vspace{1mm}
			\tabincell{l}{Carrier frequency}   &   $f_c$&   80$\mathrm{GHz}$\\
			\vspace{1mm}
			\tabincell{l}{Wavelength}   &   $\alpha$&   $3.76\mathrm{mm}$\\
			\vspace{1mm}
			\tabincell{l}{SNR} &   $\textup{SNR}$&   0$\mathrm{dB}$\\
			\vspace{1mm}
			\tabincell{l}{Angle spread in ``one-ring'' model} &   $\Psi$&   $20^\circ$\\
			\vspace{1mm}
			\tabincell{l}{Antenna number at the BS side}     &   $M$ &   32\\
			\vspace{1mm}
			\tabincell{l}{Relative distance of antennas $m1$, $m2$}      &   $D_{m1,m2}$ &   $7.52(m1-m2)\mathrm{mm}$\\
			\vspace{1mm}
			\tabincell{l}{Pilot sequence length}      &   $T$ &   32\\
			\hline
		\end{tabular}
	\end{table}

	\begin{figure}
		\setlength{\abovecaptionskip}{-0.1 cm}
		\centering
		\includegraphics[width=9.5 cm]{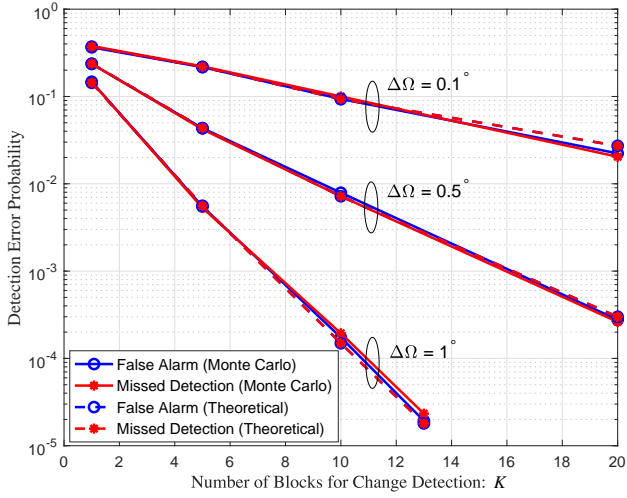}
		\caption{False alarm and missed detection probabilities under the genie-aided change detector.}
		\label{Result1}
		\vspace{-0.3cm}
	\end{figure}

\begin{figure}
		\setlength{\abovecaptionskip}{-0.1 cm}
		\centering
		\includegraphics[width=9.5 cm]{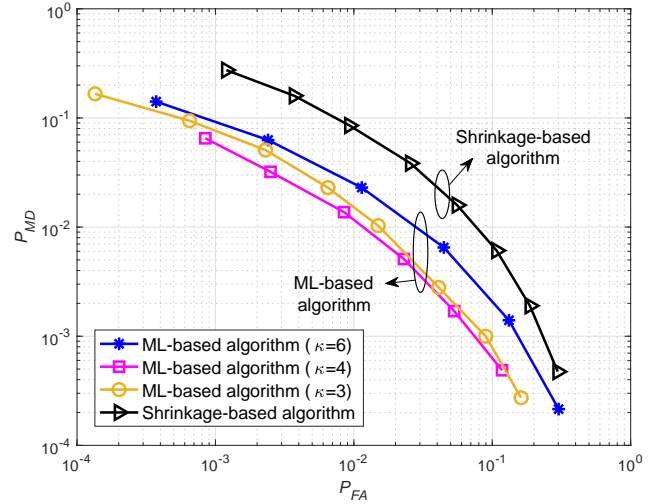}
		\caption{False alarm and missed detection probabilities with unknown $\boldsymbol{C}_1^h$ when $K=30$ and $\Delta\Omega=0.75^\circ$.}
		\label{Result2}
		\vspace{-0.4cm}
	\end{figure}

	First, we illustrate the performance of the genie-aided change estimator when $\boldsymbol{C}_1^h$ is known. Fig. \ref{Result1} shows the detection error probability as given in (\ref{DetectErr}) versus the number of detection coherence blocks when $\Delta\Omega=0.1^\circ,0.5^\circ,1^\circ$, respectively. Particularly, we select the threshold $\Theta$ such that the probabilities of false alarm and missed detection are the same. First, it is observed that the error probability obtained via Monte Carlo simulation matches perfectly with that from Theorem \ref{thm}. Next, it is observed that when $\boldsymbol{C}_1^h$ is known, very good change detection accuracy can be obtained even when $K$ is much smaller than $M$. Last, it is observed that the detection accuracy increases when the change in the channel covariance matrix is more significant.
	
	Next, we consider the case when $\boldsymbol{C}_1^h$ is unknown. We adopt the following benchmark scheme.

\textit{Benchmark Scheme (Shrinkage-based Estimation):} Based on the shrinkage technique in \cite{5484583}, the estimation of $\boldsymbol{C}^h_1$ can be expressed as
	\begin{equation}
		\hat{\boldsymbol{C}}^h_{\textup{SH}}=\Big((1-\rho)\boldsymbol{S}_{\tilde{H}}+\rho\frac{\textup{tr}(\boldsymbol{S}_{\tilde{H}})}{M}\boldsymbol{I}\Big)-\frac{\sigma^2}{E_0}\boldsymbol{I},
	\end{equation}
	where $0\leq\rho\leq1$ is the shrinkage parameter to define the weight of the sample covariance matrix in the estimation. According to \cite{5484583}, a good choice of $\rho$ is
	\begin{equation}
		\rho^* = \min\Bigg(\frac{-\frac{1}{M}\textup{tr}(\boldsymbol{S}_{\tilde{H}}\boldsymbol{S}_{\tilde{H}})+\textup{tr}^2(\boldsymbol{S}_{\tilde{H}})}{\frac{ K-1}{M}\big(\textup{tr}(\boldsymbol{S}_{\tilde{H}}\boldsymbol{S}_{\tilde{H}})-\frac{\textup{tr}^2(\boldsymbol{S}_{\tilde{H}})}{M}\big)},1\Bigg).
	\end{equation}
	Based on (\ref{criterion}), the change detector under the shrinkage estimation of $\boldsymbol{C}_1^h$ is given by
	\begin{equation}
		S(\tilde{\boldsymbol{H}},\boldsymbol{C}^h_0,\hat{\boldsymbol{C}}_{\textup{SH}}^h)\mathop{\lessgtr}^{H_0}_{H_1} \Theta.
		\label{Criterion_SH}
	\end{equation}

Fig. \ref{Result2} shows the performance comparison between the change detectors based on the ML estimation and shrinkage-based estimation of $\boldsymbol{C}_1^h$ when $K=30<M$ and $\Delta\Omega=0.75 ^\circ$. By varying the threshold $\Theta$, we can get the whole trade-off between the false alarm probability and missed detection probability. Firstly, it is observed that even if $\boldsymbol{C}_1^h$ is unknown, the error probability of our proposed change detectors is quite low. Secondly, it is observed that changed detector based on the ML estimation performs better than the change detector based on the shrinkaged-based estimation when $K<M$. Thirdly, it is observed that the change detector based on the ML-based estimation with $\kappa=4$ performs better than that with $\kappa=6$, while performs worse than that with $\kappa=3$, which indicates that an appropriate constraint on the condition number of $\hat{\boldsymbol{C}}_h^1$ is demanded to guarantee high performance of the proposed ML estimation.

\begin{figure}[t]
		\setlength{\abovecaptionskip}{-0.1 cm}
		\centering
		\includegraphics[width=9.5 cm]{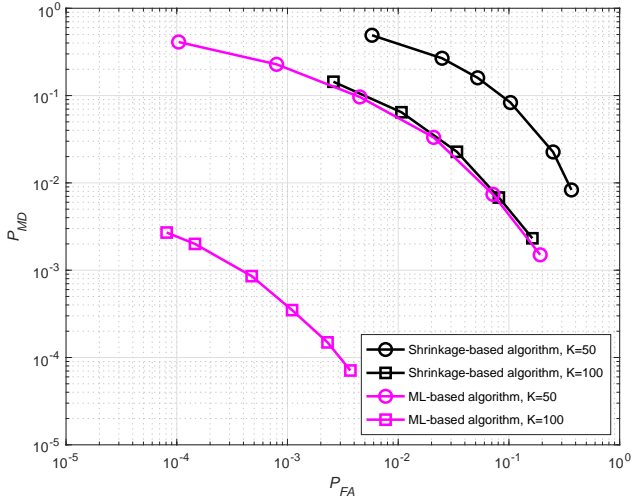}
		\caption{False alarm and missed detection probabilities with unknown $\boldsymbol{C}_1^h$ when $K=50,100$ and $\Delta\Omega=0.5^\circ$.}
		\label{Result3}
		\vspace{-0.3cm}
	\end{figure}

	We further study the performance of the changed detectors based on the ML estimation and the shrinkage estimation when $K>M$. Fig. \ref{Result3} shows the performance comparison of these two change detectors when $\Delta\Omega$ is reduced from $0.75 ^\circ$ to $0.5 ^\circ$ such that events of false alarm and missed detection can be observed in our Monte Carlo simulation with larger $K$. It is observed that when $K$ increases from $50$ to $100$, the performance gap between the ML estimation based change detector and the shrinkage estimation based change detector becomes larger. However, as $K$ becomes larger and larger, the performance gap will vanish because at last, both the ML estimation and the shrinkage estimation will lead to very accurate estimation when $K$ is sufficiently large. We do not show this result since when $K$ is very large, the error probability is very low and no false alarm and missed detection events are observed in our Monte Carlo simulation.

	\section{Conclusions}
In this paper, we studied channel covariance change detection in a system consisting of a multi-antenna BS and a single-antenna UE. First, we proposed a genie-aided change detector based on the LLR test, assuming that the knowledge of the new channel covariance matrix is known. The probabilities of false alarm and missed detection were characterized, which may serve as performance limit. Then, we considered the practical scenario with unknown new channel covariance matrix and proposed an LLR-based change detector based on the ML estimation technique. Numerical results that when the new channel covariance matrix is unknown, our proposed strategies can detect the change very accurately even with a small number of samples.
	
	\begin{appendix}
	\subsection{Proof of Theorem \ref{thm}}\label{appendix1}
	We first analyze the probability of missed detection $P_{MD}$ in (\ref{DetectErr}), which follows that
	\begin{equation}
		\begin{aligned}
			&P\big(S(\tilde{\boldsymbol{H}},\boldsymbol{C}^h_0,\boldsymbol{C}^h_1)\leq\Theta|H_1\big)\\
			=&P\Big(K\big(R+\mathrm{tr}(\boldsymbol{M}\boldsymbol{S}_{\tilde{H}})\big)\leq\Theta\Big)\\
			=&P\big\{\sum_{k=1}^{K} \tilde{\boldsymbol{h}}^H_k\boldsymbol{M}\tilde{\boldsymbol{h}}_k\leq\Theta-KR\big\}\\
			=&P\big\{2\sum_{k=1}^{K} \tilde{\boldsymbol{h}}^H_k\boldsymbol{M}\tilde{\boldsymbol{h}}_k\leq2(\Theta-KR)\big\}
		\end{aligned}
		\label{PMD}
	\end{equation}
	where $R$ and $\boldsymbol{M}$ are defined in (\ref{R}) and (\ref{Mdef}), respectively.
	
	As shown in \cite{das2020method},  $2\sum_{k=1}^{K}\tilde{\boldsymbol{h}}^H_k\boldsymbol{M}\tilde{\boldsymbol{h}}_k$ can be expressed as
	\begin{equation}
		\begin{aligned}
			2\sum_{k=1}^{K}\tilde{\boldsymbol{h}}^H_k\boldsymbol{M}\tilde{\boldsymbol{h}}_k=&K\sum_{m=1}^{M} q_{1,m}\chi^2_{2,m}\\
			=&\sum_{m=1}^{M} q_{1,m}\chi^2_{2K,m}\\
			=&\xi(\boldsymbol{q}_1,K,M),
		\end{aligned}
		\label{GenalizedChi}
	\end{equation}
	where $q_{1,m}$ and $\chi^2_{2K,m}$ are defined in Theorem \ref{thm}. From (\ref{GenalizedChi}), $2\sum_{k=1}^{K}\tilde{\boldsymbol{h}}^H_k\boldsymbol{M}\tilde{\boldsymbol{h}}_k$ is expressed as a weighted sum of chi-square variable $\chi^2_{2K,m}$, thus following the generalized chi-squared distribution. Therefore, $P_{MD}$ is equal to the CDF of $\xi(\boldsymbol{q}_1,K,M)$ at $2(\Theta-KR)$. Similarly, probability of false alarm $P_{FA}$ can be expressed as CDF of $\xi(\boldsymbol{q}_0,K,M)$ at $-2(\Theta-KR)$. Theorem \ref{thm} is thus proved.

\end{appendix}

	\bibliographystyle{IEEEtran}
	\bibliography{ChgDet}
	
\end{document}